\documentclass[12pt,english]{article}
\usepackage[T1]{fontenc}
\usepackage[latin9]{inputenc}
\usepackage[a4paper]{geometry}
\usepackage{amsmath}
\usepackage{setspace}
\usepackage{amssymb}
\makeatletter
\newcommand{\lyxaddress}[1]{
\par {\raggedright #1
\vspace{1.4em}
\noindent\par}
}

\makeatother

\usepackage{babel}

\begin{document}

\title{Octonionic Reformulation of Vector Analysis}

\author{Bhupendra C. S. Chauhan, P. S. Bisht and O. P. S. Negi%
\thanks{Present Address from November 08- December22, 2010:\textbf{ Universität
Konstanz, Fachbereich Physik, Postfach-M 677, D-78457 Konstanz, Germany}%
} }

\maketitle

\lyxaddress{\begin{center}
Department of Physics\\
Kumaun University\\
 S. S. J. Campus\\
Almora \textendash{} 263601 (Uttarakhand)
\par\end{center}}

\lyxaddress{\begin{center}
Email: bupendra.123@gmail.com \\
ps\_bisht123@rediffmail.com \\
ops\_negi@yhaoo.co.in
\par\end{center}}
\begin{abstract}
According to celebrated Hurwitz theorem, there exists four division
algebras consisting of R (real numbers), C (complex numbers), H (quaternions)
and O (octonions). Keeping in view the utility of octonion variable
we have tried to extend the three dimensional vector analysis to seven
dimensional one. Starting with the scalar and vector product in seven
dimensions, we have redefined the gradient, divergence and curl in
seven dimension. It is shown that the identity $n(n-1)(n-3)(n-7)=0$
is satisfied only for $0$, $1$, $3$ and $7$ dimensional vectors.
We have tried to write all the vector inequalities and formulas in
terms of seven dimensions and it is shown that same formulas loose
their meaning in seven dimensions due to non-associativity of octonions.
The vector formulas are retained only if we put certain restrictions
on octonions and split octonions.
\end{abstract}

\section{Introduction}

Octonions were first introduced in Physics by Jordan, Von Neuman and
Wigner \cite{key-1}, who investigated a new finite Hilbert space,
on replacing the complex numbers by Octonions \cite{key-2}. According
to celebrated Hurwitz theorem \cite{key-3} there exits four-division
algebra consisting of $\mathbb{R}$ (real numbers), $\mathbb{C}$
(complex numbers), $\mathbb{H}$ (quaternions) and $\mathcal{O}$
(octonions). All four algebras are alternative with antisymmetric
associators. Real numbers and complex numbers are limited only up
to two dimensions, quaternions are extended to four dimensions (one
real and three imaginaries) while octonions represent eight dimensions
(one scalar and seven vectors namely one real and seven imaginaries).
In 1961, Pais \cite{key-4} pointed out a striking similarity between
the algebra of interactions and the split octonion algebra. Accordingly,
some attention has been drawn to octonions in theoretical physics
with the hope of extending the 3+1 space-time framework of the theory
to eight dimensions in order to accommodate the ever increasing quantum
numbers and internal symmetries assigned to elementary particles and
gauge fields. A lot of literature is available \cite{key-5,key-6,key-7,key-8,key-9,key-10,key-11,key-12}
on the applications of octonions to interpret wave equation, Dirac
equation, and the extension of octonion non - associativity to physical
theories. Keeping in view the utility of octonion variable, in the
present paper, we have tried to extend the three dimensional vector
analysis to seven dimensional one. Starting with the scalar and vector
product, we have redefined the gradient, divergence and curl in seven
dimension with the definitions of octonion variables. It is shown
that the identity $n(n-1)(n-3)(n-7)=0$ is satisfied only for $0,1,3$
and $7$ dimensional vectors. We have tried to write all the vector
inequalities and formulas in terms of seven dimensions and it is shown
that same formulas loose their meaning in seven dimensions due to
non \textendash{} associativity of octonions. In this context we have
tried to reformulate the vector analysis and it is shown that some
vector identities are loosing their original form of three dimensional
space in seven dimension vector space and looking for more generalized
representation in seven dimension space.

\section{Octonion Definition}

An octonion $x$ is expressed as a set of eight real numbers

\begin{align}
x= & e_{0}x_{0}+e_{1}x_{1}+e_{2}x_{2}+e_{3}x_{3}+e_{4}x_{4}+e_{5}x_{5}+e_{6}x_{6}+e_{7}x_{7}\nonumber \\
= & e_{0}x_{0}+\sum_{j=1}^{7}e_{j}x_{j}\label{eq:1}\end{align}
where $e_{j}(\forall\, j=1,2,...,7)$ are imaginary octonion units
and $e_{0}$ is the multiplicative unit element. Set of octets $(e_{0},\, e_{1},\, e_{2},e_{3},e_{4},e_{5},e_{6},e_{7})$
are known as the octonion basis elements and satisfy the following
multiplication rules

\begin{eqnarray}
e_{0}=1;\,\, & e_{0}e_{j}=e_{j}e_{0}=e_{j}; & \,\,\,\, e_{j}e_{k}=-\delta_{jk}e_{0}+f_{jkl}e_{l}.\,\,(\forall j,k,l=1,2,.....,7)\label{eq:2}\end{eqnarray}
The structure constants $f_{jkl}$ is completely antisymmetric and
takes the value $1$ for following combinations,

\begin{equation}
f_{jkl}=+1\left\{ \forall(jkl)=(123),\,(471),\,(257),\,(165),\,(624),\,(543),\,(736)\right\} .\label{eq:3}\end{equation}
It is to be noted that the summation convention is used for repeated
indices. Here the octonion algebra $\mathcal{O}$ is described over
the algebra of real numbers having the vector space of dimension $8$.
Octonion conjugate is defined as

\begin{align}
\overline{x} & =e_{0}x_{0}-e_{1}x_{1}-e_{2}x_{2}-e_{3}x_{3}-e_{4}x_{4}-e_{5}x_{5}-e_{6}x_{6}-e_{7}x_{7}\nonumber \\
= & e_{0}x_{0}-\sum_{j=1}^{7}e_{j}x_{j}\label{eq:4}\end{align}
where we have used the conjugates of basis elements as $\overline{e_{0}}=e_{0}$
and $\overline{e_{A}}=-e_{A}$. Hence an octonion can be decomposed
in terms of its scalar $(Sc(x))$ and vector $(Vec(x))$ parts as 

\begin{eqnarray}
Sc(x) & = & \frac{1}{2}(\, x\,+\,\overline{x}\,);\,\,\,\,\,\, Vec(x)=\frac{1}{2}(\, x\,-\,\overline{x}\,)=\sum_{j=1}^{7}\, e_{j}x_{j}.\label{eq:5}\end{eqnarray}
Conjugates of product of two octonions is described as $\overline{(x\, y)}=\overline{y}\,\,\overline{x}$
while the own conjugate of an octonion is written as $\overline{(\overline{x})}=x$.
The scalar product of two octonions is defined as 

\begin{eqnarray}
\left\langle x\,,\, y\right\rangle  & =\frac{1}{2}(x\,\overline{y}+y\,\overline{x})=\frac{1}{2}(\overline{x}\, y+\overline{y}\, x)= & \sum_{\alpha=0}^{7}\, x_{\alpha}\, y_{\alpha}.\label{eq:6}\end{eqnarray}
The norm $N(x)$ and inverse $x^{-1}$(for a nonzero $x$) of an octonion
are respectively defined as

\begin{eqnarray}
N(x)=x\,\overline{x}=\overline{x}\, x & = & \sum_{\alpha=0}^{7}\, x_{\alpha}^{2}.e_{0};\nonumber \\
x^{-1} & = & \frac{\overline{x}}{N(x)}\,\Longrightarrow x\, x^{-1}=x^{-1}\, x=1.\label{eq:7}\end{eqnarray}
The norm $N(x)$ of an octonion $x$ is zero if $x=0$, and is always
positive otherwise. It also satisfies the following property of normed
algebra

\begin{eqnarray}
N(x\, y)= & N(x)\, N(y)= & N(y)\, N(x).\label{eq:8}\end{eqnarray}
Equation (\ref{eq:2}) directly leads to the conclusion that octonions
are not associative in nature and thus do not form the group in their
usual form. Non - associativity of octonion algebra $\mathcal{O}$
is described by the associator $(x,y,z)=(xy)z-x(yz)\,\,\forall x,y,z\in\mathcal{O}$
defined for any $3$ octonions. If the associator is totally antisymmetric
for exchanges of any $2$ variables, i.e. $(x,y,z)=-(z,y,x)=-(y,x,z)=-(x,z,y)$,
the algebra is called alternative. Hence, the octonion algebra is
neither commutative nor associative but, is alternative.

\section{Multi- Dimensional Vector Analysis}

Following Silagadze \cite{key-11}, let us consider $n$-dimensional
vector space $\mathbb{R}^{n}$over the field of real numbers with
standard Euclidean scalar product. So, it is natural to describe the
generalization of usual $3$- dimensional vector space to $n$-dimensional
vector space in order to represent the following three dimensional
vector products of two vectors in $n$-dimensional vector space i.e. 

\begin{align}
\overrightarrow{A}\times\overrightarrow{A} & =0,\label{eq:9}\\
\left(\overrightarrow{A}\times\overrightarrow{B}\right)\centerdot\overrightarrow{A} & =\left(\overrightarrow{A}\times\overrightarrow{B}\right)\centerdot\overrightarrow{B}=0,\label{eq:10}\\
\left|\overrightarrow{A}\times\overrightarrow{B}\right|= & \left|\overrightarrow{A}\right|\left|\overrightarrow{B}\right|,\,\,\,\, if\,\,\,\,\left(\overrightarrow{A}\centerdot\overrightarrow{B}\right)=0,\label{eq:11}\\
\left(\overrightarrow{A}\times\overrightarrow{B}\right)\centerdot\left(\overrightarrow{A}\times\overrightarrow{B}\right) & =\left(\overrightarrow{A}\centerdot\overrightarrow{A}\right)\left(\overrightarrow{B}\centerdot\overrightarrow{B}\right)-\left(\overrightarrow{A}\centerdot\overrightarrow{B}\right)^{2},\label{eq:12}\\
\overrightarrow{A}\times\left(\overrightarrow{B}\times\overrightarrow{A}\right)= & \left(\overrightarrow{A}\centerdot\overrightarrow{A}\right)\overrightarrow{B}-\left(\overrightarrow{A}\centerdot\overrightarrow{B}\right)\overrightarrow{A},\label{eq:13}\end{align}
where$\overrightarrow{A}$ and $\overrightarrow{B}$ are the vectors
in $n$-dimensional vector space. However the familiar identity 

\begin{align}
\overrightarrow{A}\times\left(\overrightarrow{B}\times\overrightarrow{C}\right)= & \overrightarrow{B}\left(\overrightarrow{A}\centerdot\overrightarrow{C}\right)-\overrightarrow{C}\left(\overrightarrow{A}\centerdot\overrightarrow{B}\right)\label{eq:14}\end{align}
is not satisfied in general for all values of $n$- dimensional vector
space.This identity (\ref{eq:14}) is verified only for the space
dimension $n$ satisfying \cite{key-11} the equation

\begin{align}
n(n-1)(n-3)(n-7) & =0.\label{eq:15}\end{align}
As such, the space dimension must be equal to magic number seven for
the unique generalization of ordinary three dimensional vector products.
It shows that the identity (\ref{eq:15}) is satisfied only for $n=$$0$,
$1$, $3$ and $7$ dimensional vectors. The identity (\ref{eq:15})
is thus the direct consequence of celebrated Hurwitz theorem \cite{key-3}
which shows that there exits four-division algebra consisting of $\mathbb{R}$
(real numbers) $(n=0)$, $\mathbb{C}$ (complex numbers) $(n=1)$,
$\mathbb{H}$ (quaternions) $(n=3)$ and $\mathcal{O}$ (octonions)
$(n=7)$.

\section{Octonion Analysis of Vector Space}

Using the octonion multiplication rules (\ref{eq:2}), we may define
the seven dimensional vector product as 

\begin{align}
\overrightarrow{e_{j}}\times\overrightarrow{e_{k}} & =\sum_{k=1}^{k=7}f_{jkl}\overrightarrow{e_{k}},\,\,\,\,\,\,\forall j,k,l=1,2,......,6,7\label{eq:16}\end{align}
where $f_{jkl}$ is described by equation (\ref{eq:3}). It is a totally
$G_{2}-$invariant anti-symmetric tensor. As such we have

\begin{align}
f_{jkl}f_{lmn}=g_{jkmn}+\delta_{jm}\delta_{kn}-\delta_{jn}\delta_{km}= & -f_{lmn}f_{jkl}\label{eq:17}\end{align}
where $g_{jkmn}=\overrightarrow{e_{j}}\centerdot\left\{ \overrightarrow{e_{k}},\overrightarrow{e_{m}},\overrightarrow{e_{n}}\right\} $
is a totally $G_{2}-$invariant anti-symmetric tensor \cite{key-11}
and $g_{jkmn}=-g_{mkjn}.$ The only independent components are $g_{1254}=g_{1267}=g_{1364}=g_{1375}=g_{2347}=g_{2365}=g_{4576}=1$.
Thus, we may write the the left hand side of equation (\ref{eq:14})
as 

\begin{align}
\vec{A}\times(\vec{B}\times\vec{C})= & -\sum_{jk=1}^{7}\sum_{pq=1}^{7}\left[g_{pqjk}+\delta_{pj}\delta_{qk}-\delta_{pk}\delta_{qj}\right]A_{j}B_{p}C_{q}\hat{e_{k}}\nonumber \\
= & -\sum_{jk}^{7}\sum_{pq}^{7}g_{pqjk}A_{j}B_{p}C_{q}\hat{e_{k}}-\sum_{jk}^{7}A_{j}B_{j}C_{k}\hat{e_{k}}+\sum_{jk}^{7}A_{j}C_{j}B_{k}\hat{e_{k}}\nonumber \\
= & irreducible\, term+\vec{B}(\vec{A}.\vec{C})-\vec{C}(\vec{A}.\vec{B}).\label{eq:18}\end{align}
Hence, the identity of Triple Product in seven dimensional vector
space is not satisfied unless we apply the definition of octonion
to define the irreducible term as a ternary product such that

\begin{align}
irreducible\, term=-\sum_{jk}^{7}\sum_{pq}^{7}g_{pqjk}A_{j}B_{p}C_{q}\hat{e_{k}}= & \left\{ \vec{A},\vec{B},\vec{C}\right\} ,\label{eq:19}\end{align}
which gives rise to 

\begin{align}
\vec{A}\times(\vec{B}\times\vec{C})= & \vec{B}(\vec{A}.\vec{C})-\vec{C}(\vec{A}.\vec{B})+\left\{ \vec{A},\vec{B},\vec{C}\right\} .\label{eq:20}\end{align}
The $\left\{ \vec{A},\vec{B},\vec{C}\right\} $ is called as the associator
for the case of octonions. Applying th condition of alternativity
to octonions the equation (\ref{eq:20}) reduces to the well known
vector identity (\ref{eq:14}) for the various permutation values
of structure constant $f_{jkl}$ for which the associator is going
to be vanished.

Let us use the vector calculus for seven dimensional vector space
from the definition of octonion variables. Defining the seven dimensional
differential operator $Nabla$ as $\vec{\nabla}=\sum_{i}^{7}\hat{e_{i}}\frac{\partial}{\partial x_{i}}$,
we may now define the gradient, curl and divergence of scar and vector
quantities as 

\begin{align}
grad\, u=\vec{\nabla}u= & \sum_{i}^{7}\hat{e_{i}}\frac{\partial u}{\partial x_{i}},\label{eq:21}\\
curl\,\overrightarrow{A}=\vec{\nabla}\times\vec{A}= & \sum_{ijk}^{7}f_{ijk}\frac{\partial A_{j}}{\partial x_{i}}\hat{e_{k}},\label{eq:22}\\
div\,\overrightarrow{A}=\vec{\nabla}.\vec{A}= & \sum_{i}^{7}\frac{\partial A_{i}}{\partial x_{i}}.\label{eq:23}\end{align}
The Divergent of a curl is zero in usual three dimensional vector-space.
This is also applicable to seven dimensional vector-space if we adopt
the octonion multiplication rules (\ref{eq:2}). We may also prove
it in the following manner as

\begin{align}
\vec{\nabla}.(\vec{\nabla}\times\vec{A})= & \sum_{l}^{7}\left[\hat{e_{l}}\frac{\partial}{\partial x_{l}}\right]\centerdot\left[\sum_{ijk}^{7}f_{ijk}\frac{\partial A_{j}}{\partial x_{i}}\hat{e_{k}}\right]=\sum_{l}^{7}\sum_{ijk}^{7}f_{ijk}\frac{\partial^{2}A_{j}}{\partial x_{l}\partial x_{i}}\delta_{lk}\nonumber \\
= & \sum_{ijk}^{7}f_{ijk}\frac{\partial^{2}A_{j}}{\partial x_{k}\partial x_{i}}=\frac{1}{2}\sum_{ijk}^{7}\left[f_{ijk}+f_{kji}\right]\frac{\partial^{2}A_{j}}{\partial x_{i}\partial x_{k}}=0.\label{eq:24}\end{align}
Hence curl of a vector is solenoidal in seven dimensional vector space.
We know that the curl of gradient of a vector is zero in $3$-dimensional
vector-space. thus we see that it also happens for the case of $7$-
dimensional vector- space by adopting the octonion multiplication
rules (\ref{eq:2} ) in the following way as 

\begin{align}
\vec{\nabla}\times(\vec{\nabla}u)= & \sum_{ijk}^{7}f_{ijk}\frac{\partial}{\partial x_{i}}\left(\sum_{j}^{7}\frac{\partial u}{\partial x_{j}}\right)\hat{e_{k}}\nonumber \\
=\sum_{ijk}^{7}f_{ijk}\frac{\partial^{2}u}{\partial x_{i}\partial x_{j}}\hat{e_{k}} & =\frac{1}{2}\sum_{ijk}^{7}[f_{ijk}+f_{jik}]\frac{\partial^{2}u}{\partial x_{j}\partial x_{i}}\hat{e_{k}}=0.\label{eq:25}\end{align}
Thus, the gradient of a curl is also irrotational in seven dimensional
vector space with the definition of octonions. Let us see what happens
to the vector identities 

\begin{align}
\vec{\nabla}\centerdot(\vec{A}\times\vec{B})= & \vec{B}\centerdot(\vec{\nabla}\times\vec{A})-\vec{A}\centerdot(\vec{\nabla}\times\vec{B})\label{eq:26}\end{align}
in seven dimensional vector-space. The left hand side of equation
(\ref{eq:26}) reduces to 

\begin{align}
\vec{\nabla}\centerdot(\vec{A}\times\vec{B})= & \sum_{l}^{7}\hat{e_{l}}\frac{\partial}{\partial x_{l}}\centerdot\left[\sum_{ijk}^{7}f_{ijk}A_{i}B_{j}\hat{e_{k}}\right]\nonumber \\
= & \sum_{ijk}^{7}\sum_{l}^{7}f_{ijk}\frac{\partial}{\partial x_{l}}[A_{i}B_{j}]\delta_{ij}=\sum_{ijk}^{7}f_{ijk}\left[A_{i}\frac{\partial B_{j}}{\partial x_{k}}+B_{j}\frac{\partial A_{i}}{\partial x_{k}}\right]\label{eq:27}\end{align}
while the right hand side of equation (\ref{eq:26}) changes to 

\begin{align}
\vec{B}\centerdot(\vec{\nabla}\times\vec{A})-\vec{A}\centerdot(\vec{\nabla}\times\vec{B})= & \sum_{l}^{7}\hat{e_{l}}B_{l}.\left[\sum_{mnt}^{7}f_{mnt}\frac{\partial A_{n}}{\partial x_{m}}\hat{e_{t}}\right]-\sum_{s}^{7}\hat{e_{s}}A_{s}.\left[\sum_{pqr}^{7}f_{pqr}\frac{\partial B_{q}}{\partial x_{p}}\hat{e_{r}}\right]\nonumber \\
= & \sum_{ijk}^{7}f_{ijk}B_{j}\frac{\partial A_{i}}{\partial x_{k}}-\left[-\sum_{ijk}^{7}f_{ijk}A_{i}\frac{\partial B_{j}}{\partial x_{k}}\right]=\sum_{ijk}^{7}f_{ijk}\left[A_{i}\frac{\partial B_{j}}{\partial x_{k}}+B_{j}\frac{\partial A_{i}}{\partial x_{k}}\right],\label{eq:28}\end{align}
which is equal to the left hand side equation (\ref{eq:27}). Hence
the identity (\ref{eq:26}) has been satisfied for seven dimensional
space. Similarly, on using the octonion multiplication rules (\ref{eq:2})
we may prove the following identities in seven dimensions i.e.

\begin{align}
\vec{\nabla\centerdot}(u\vec{A})= & u(\vec{\nabla}\centerdot\vec{A})+\vec{A}\centerdot(\vec{\nabla}u),\label{eq:29}\\
\vec{\nabla}\times(u\vec{A})= & u(\vec{\nabla}\times\vec{A})+(\vec{\nabla}u)\times\vec{A},\label{eq:30}\\
\vec{\nabla}\times(\vec{\nabla}\times\vec{A}) & =\vec{\nabla}(\vec{\nabla}\centerdot\vec{A})-\nabla^{2}\vec{A}.\label{eq:31}\end{align}
However the vector identity

\begin{align}
\vec{\nabla}\times(\vec{A}\times\vec{B})= & (\vec{B}\centerdot\vec{\nabla})\vec{A}-(\vec{A}\centerdot\vec{\nabla})\vec{B}-\vec{B}(\vec{\nabla}\centerdot\vec{A})+\vec{A}(\vec{\nabla}\centerdot\vec{B})\label{eq:32}\end{align}
is not satisfied and its left hand side reduces to 

\begin{align}
\vec{\nabla}\times(\vec{A}\times\vec{B})= & \sum_{wkt}^{7}f_{wkt}\frac{\partial}{\partial x_{w}}\left[\sum_{ijk}^{7}f_{ijk}A_{i}B_{j}\right]\hat{e_{t}}=-\sum_{wkt}^{7}\sum_{ijk}^{7}f_{ijk}f_{kwt}\frac{\partial}{\partial x_{w}}[A_{i}B_{j}]\hat{e_{t}}\label{eq:33}\end{align}
whereas the right hand expression of equation (\ref{eq:32}) becomes

\begin{equation}
(\vec{B}\centerdot\vec{\nabla})\vec{A}-(\vec{A}\centerdot\vec{\nabla})\vec{B}-\vec{B}(\vec{\nabla}\centerdot\vec{A})+\vec{A}(\vec{\nabla}\centerdot\vec{B})=\sum_{ij}^{7}\frac{\partial A_{i}}{\partial x_{j}}B_{j}\hat{e_{i}}-\sum_{lm}^{7}A_{l}\frac{\partial B_{m}}{\partial x_{l}}\hat{e_{m}}-\sum_{ij}^{7}\frac{\partial A_{i}}{\partial x_{i}}B_{j}\hat{e_{j}}+\sum_{ij}^{7}A_{i}\frac{\partial B_{j}}{\partial x_{j}}\hat{e_{i}}.\label{eq:34}\end{equation}
Comparing equations (\ref{eq:33}) and (\ref{eq:34}), we get 

\begin{align}
\vec{\nabla}\times(\vec{A}\times\vec{B})= & (\vec{B}\centerdot\vec{\nabla})\vec{A}-(\vec{A}\centerdot\vec{\nabla})\vec{B}-\vec{B}(\vec{\nabla}\centerdot\vec{A})+\vec{A}(\vec{\nabla}\centerdot\vec{B})+irreducible\, term\label{eq:35}\end{align}
where the irreducible term takes the expression 

\begin{align}
irreducible\, term= & -\sum_{wt}^{7}\sum_{ij}^{7}g_{ijwt}\frac{\partial}{\partial x_{w}}[A_{i}B_{j}]\hat{e_{t}}\nonumber \\
=-\sum_{wt}^{7}\sum_{ij}^{7}g_{ijwt}\frac{\partial}{\partial x_{w}}[A_{i}B_{j}]\hat{e_{t}} & =\sum_{t}^{7}\hat{e_{t}}\left[\vec{\nabla}\centerdot\left\{ \hat{e_{t}},\vec{A,}\vec{B}\right\} \right].\label{eq:36}\end{align}
Hence we get \begin{align}
\vec{\nabla}\times(\vec{A}\times\vec{B})+\sum_{t}^{7}\hat{e_{t}}\left[\vec{\nabla}\centerdot\left\{ \hat{e_{t}},\vec{A,}\vec{B}\right\} \right] & =(\vec{B}\centerdot\vec{\nabla})\vec{A}-(\vec{A}\centerdot\vec{\nabla})\vec{B}-\vec{B}(\vec{\nabla}\centerdot\vec{A})+\vec{A}(\vec{\nabla}\centerdot\vec{B}).\label{eq:37}\end{align}
Thus we see that the identity (\ref{eq:32}) is satisfied only when
we have the reducible term vanishing and that can be obtained for
different values of permutations of structure constant given by equation
(\ref{eq:3}). 

Similarly the other vector identity 

\begin{align}
\vec{A}\times(\vec{\nabla}\times\vec{B})+\vec{B}\times(\vec{\nabla}\times\vec{A})+(\vec{A}\centerdot\vec{\nabla})\vec{B}+\vec{B}(\vec{\nabla}\centerdot\vec{A})= & \vec{\nabla}(\vec{A}\centerdot\vec{B})\label{eq:38}\end{align}
may also be verified after doing certain modifications also. For this
, let us take the first and third terms of left hand side of equation
(\ref{eq:38}) as

\begin{align}
\vec{A}\times(\vec{\nabla}\times\vec{B})+(\vec{A}\centerdot\vec{\nabla})\vec{B} & =\sum_{ikst}^{7}g_{stik}A_{i}\frac{\partial B_{t}}{\partial x_{s}}\hat{e_{k}}+\sum_{ik}^{7}A_{i}\frac{\partial B_{i}}{\partial x_{k}}\hat{e_{k}}\label{eq:39}\end{align}
and the other two terms of left hand side of equation (\ref{eq:38})
are written as 

\begin{align}
\vec{B}\times(\vec{\nabla}\times\vec{A})+\vec{B}(\vec{\nabla}\centerdot\vec{A}) & =\sum_{ikst}^{7}g_{stik}B_{i}\frac{\partial A_{t}}{\partial x_{s}}\hat{e_{k}}+\sum_{ik}^{7}B_{i}\frac{\partial A_{i}}{\partial x_{k}}\hat{e_{k}}\label{eq:40}\end{align}
However, the right hand side of (\ref{eq:38}) reduces to

\begin{align}
\vec{\nabla}(\vec{A}\centerdot\vec{B})= & \sum_{ik}^{7}A_{i}\frac{\partial B_{i}}{\partial x_{k}}\hat{e_{k}}+\sum_{ik}^{7}B_{i}\frac{\partial A_{i}}{\partial x_{k}}\hat{e_{k}}.\label{eq:41}\end{align}
As such, it is concluded that the general form of equation (\ref{eq:38})
of three dimensional vector space is not satisfied for octonions with
the fact that this equation becomes 

\begin{align}
\vec{A}\times(\vec{\nabla}\times\vec{B})+(\vec{A}\centerdot\vec{\nabla})\vec{B}+\vec{B}\times(\vec{\nabla}\times\vec{A})+(\vec{B}\centerdot\vec{\nabla})\vec{A}= & \vec{\nabla}(\vec{A}\centerdot\vec{B})+irreducible\, term.\label{eq:42}\end{align}
Applying the alternativity relations for octonion basis elements,
we see that the the irreducible term is reduced as 

\begin{align}
irreducible\, term=- & \left[\sum_{ikst}^{7}g_{stik}A_{i}\frac{\partial B_{t}}{\partial x_{s}}\hat{e_{k}}+\sum_{ikst}^{7}g_{stik}B_{i}\frac{\partial A_{t}}{\partial x_{s}}\hat{e_{k}}\right]\nonumber \\
= & -\left[\sum_{k}^{7}\hat{e_{k}}[\vec{A}\centerdot\{\hat{e_{k}},\vec{\nabla},\vec{B}\}]+\sum_{k}^{7}\hat{e_{k}}[\vec{B}\centerdot\{\hat{e_{k}},\vec{\nabla},\vec{A}\}]\right]=0.\label{eq:43}\end{align}
Hence the identity (\ref{eq:38}) is verified for seven dimensional
vector space in terms of octonion basis elements.. 

\textbf{Acknowledgment: }One of us OPSN is thankful to Professor H.
Dehnen, Universität Konstanz, Fachbereich Physik, Postfach-M 677,
D-78457 Konstanz, Germany for his kind hospitality at Universität
Konstanz. He is also grateful to German Academic Exchange Service
(Deutscher Akademischer Austausch Dienst), Bonn for their financial
support under DAAD re-invitation programme.

\end{document}